# Coherent phonon induced optical modulation in semiconductors at terahertz frequencies

**Muneaki Hase[1], Masayuki Katsuragawa[2], Anca Monia Constantinescu[3], and Hrvoje Petek[3]**

[1]Institute of Applied Physics, University of Tsukuba, 1-1-1 Tennodai, Tsukuba 305-8573, Japan.
[2]Department of Applied Physics and Chemistry, University of Electro-Communications, 1-5-1 Chofugaoka, Chofu, 182-8585, Tokyo, Japan,
[3]Department of Physics and Astronomy, University of Pittsburgh, PA 15260, USA.
E-mail: mhase@bk.tsukuba.ac.jp

**Abstract.** The coherent modulation of electronic and vibrational nonlinearities in atoms and molecular gases by intense few-cycle pulses has been used for high-harmonic generation in the soft X-ray and attosecond regime, as well as for Raman frequency combs that span multiple octaves from the Terahertz to Petahertz frequency regions. In principle, similar high-order nonlinear processes can be excited efficiently in solids and liquids on account of their high nonlinear polarizability densities. In this paper, we demonstrate the phononic modulation of the optical index of Si and GaAs for excitation and probing near their direct band gaps, respectively at ~3.4 eV and ~3.0 eV. The large amplitude coherent longitudinal optical polarization due to the excitation of longitudinal optical (LO) phonon of Si (001) and LO phonon-plasmon coupled modes in GaAs (001) excited by 10-fs laser pulses induces effective amplitude and phase modulation of the reflected probe light. The combined action of the amplitude and phase modulation in Si and GaAs generates phonon frequency combs with more than 100 and 60 THz bandwidth, respectively.

## 1. Introduction

Advances in femtosecond laser technology enable observation of optical response from elementary excitations in solids in real time [1, 2]. In particular, there have been extensive studies focused on the excitation and dephasing of the coherent optical phonons in polar and non-polar semiconductors (GaAs [3], Ge [4], Si [5, 6]), semimetals (Bi [7, 8] and Sb[7]), metals [9, 10], ferroelectrics [11, 12], and organic crystals [13]. Because combining optical and electronic components could revolutionize semiconductor device performance and functionality [14], there is substantial interest in both to observing and controlling the coherent optical response of Si and GaAs through nonlinear optical interactions [15].

A major breakthrough in study of coherence in Si has been the observation of zone-center coherent optical phonons by Sabbah and Riffe [5]. They measured the isotropic transient reflectivity following the excitation with 28 fs duration pulses at 800 nm. The optical phonon oscillation led to modulation of the reflectivity of a Si/SiO₂ surface with an amplitude of $\Delta R/R_0$ of only $7 \times 10^{-6}$; the sine phase of coherent phonon signal implied excitation by the non-resonant impulsive stimulated Raman scattering (ISRS) [11, 13]. Subsequently, employing the second harmonic light of a Ti:sapphire oscillator (10 fs pulse duration; 406 nm central wavelength) with approximately eight times smaller excitation fluence, Hase *et al*. observed the coherent LO phonon oscillations in *n*-Si [6] with an amplitude of $\Delta R_{eo}/R_0 \sim 6 \times 10^{-5}$. The nearly two orders-of-magnitude larger signal amplitude and the approximate cosine phase implicated the displacive resonant ISRS mechanism [16].

Among compound semiconductors, GaAs is a promising material for device applications such as ultrafast optical switching [17] and generation of terahertz radiation [18]. In polar semiconductors such as Si-doped *n*-type GaAs, it is well known that plasmons and LO phonons couple through Coulomb interactions to form a pair of hybrid modes; the frequencies of these LO phonon-plasmon coupled (LOPC) modes ($\omega_\pm$)

$$\omega_\pm^2 = \frac{1}{2}\left(\omega_p^2 + \omega_{LO}^2\right) \pm \frac{1}{2}\left[\left(\omega_p^2 + \omega_{LO}^2\right)^2 - 4\omega_p^2\omega_{TO}^2\right]^{\frac{1}{2}}, \qquad (1)$$

depend on the carrier density *N* according through the plasma frequency $\omega_P = \sqrt{4\pi e^2 N / \varepsilon_\infty m^*}$ [19], where $\omega_{LO}$ and $\omega_{TO}$ are the LO and TO phonon frequencies, $\varepsilon_\infty$ is the high-frequency dielectric constant and $m^*$ is the reduced electron-hole mass. Recently, ultrafast relaxation of the coherent LOPC modes in *n*-GaAs has been studied using pump-probe techniques [20, 21]. It was demonstrated that the frequency of the LOPC modes is determined by the total electron density, including from the static impurity doping, thermal excitation, and the optical band-gap excitation. One can even deduce the carrier mobility values in GaAs from the relaxation times of the observed coherent LOPC modes [22]. Thus, the frequencies of coherent LOPC modes are sensitive to both the static- and photoexcited-carrier densities. Because the photoexcited carrier distributions evolve on the femtosecond time scale, the transient LOPC mode frequencies reflect the ultrafast non-equilibrium plasma dynamics in ionic semiconductors [23].

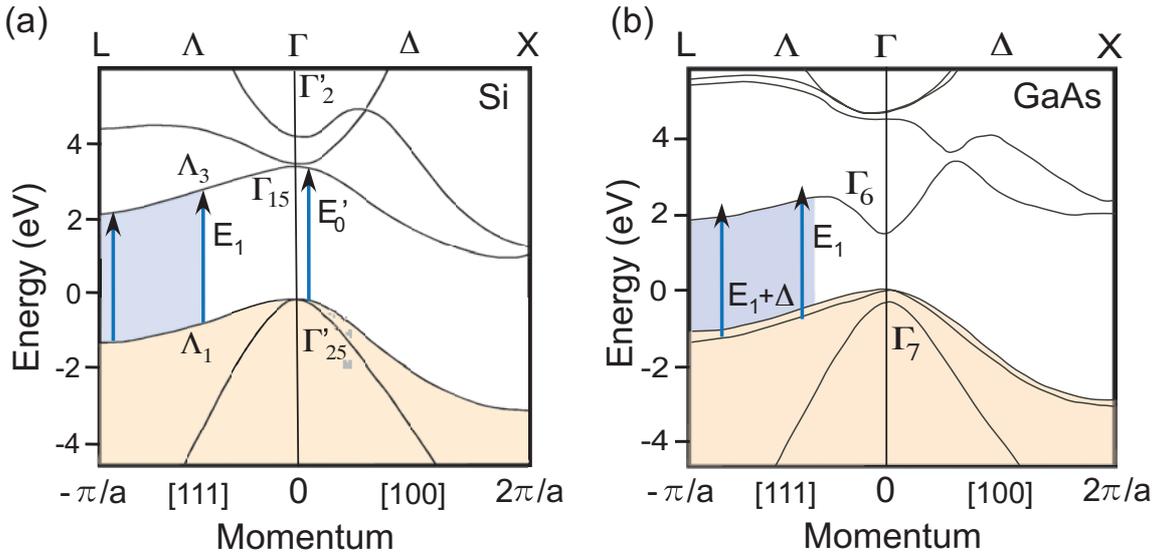

**Figure 1.** Electronic band structures for (a) Si and (b) GaAs. The vertical arrows and blue-colored regions along the Γ-L line represent optical transitions due to photo-excitation with 397 nm (central wavelength) light.

To date Raman comb generation by molecular modulation has been examined in a gas phase, where coherent molecular motion acts on the laser light to produce a wide frequency modulated spectrum [24, 25]. Moreover, high-harmonic coherent acoustic phonons have been observed in semiconductor superlattices and quantum well structures, where the attainable frequencies are characteristic of the speed of sound and the nanostructure dimensions [26, 27]. Although light modulators based on acoustic phonons can be engineered for high-harmonic operation through design of the semiconductor superlattice zone-folding, and strain, it is difficult to achieve operation in the THz frequency range. In the case of the optical phonon mode, however, the origin of high-harmonic generation can be significantly different; frequency combs of optical modes can be generated through high-order optical nonlinearity, e.g., high-order Raman scattering [28, 29], or the anharmonic crystal lattice response [12, 30].

In this paper, we present the coherent phonon induced complex refractive index modulation in Si and GaAs wafers for the excitation and probing at variable wavelength around 400 nm, which is just below the direct band gap in Si and resonant with the direct band gap in GaAs. In both cases the direct band gap excitation involves two nearly overlapping transitions. In the case of Si, E'$_0$ and E$_1$ critical points at 3.320 and 3.396 eV involve transitions at the $\Gamma$ point and for a range of momenta along the $\Lambda$–L (Fig. 1a) [31]. The E$_1$ critical points has a significantly larger transition moment, and is dominant in the resonance enhancement of Raman spectra [32]. For GaAs the E$_1$ and E$_1$ + $\Delta$ critical points with nearly equal transition moments at 3.017 and 3.245 eV couple states for a range of momenta along the $\Gamma$–L (Fig. 1b). The excitation of Si(001) with 10-fs laser pulses impulsively drives coherent optical phonons with 15.6 THz frequency, which modulate sample reflectivity and generate a broad comb of frequencies beyond 100 THz; in the case of GaAs(001), the photocarrier density dependent LOPC mode at ~7.8 THz modulates sample reflectivity and generate a comb of frequencies up to 60 THz.

## 2. Experimental: Ultrafast electro-optic sampling measurements

The anisotropic transient reflectivity of *n*-doped ($1.0 \times 10^{15}$ cm$^{-3}$) Si(001) and *n*-doped ($7.0 \times 10^{17}$ cm$^{-3}$) GaAs(001) samples were measured in air at 295 K by the electro-optic (e-o) sampling technique [3, 8]. Nearly collinear, pump and probe beams [397 nm (3.12 eV) central wavelength] were overlapped at a $7.2 \times 10^{-7}$ cm$^2$ spot on the sample. The maximum average pump power from a 70 MHz repetition rate, frequency doubled Ti:sapphire laser oscillator of 60 mW (1.0 mJ/cm$^2$) generated $N \approx 1.0 \times 10^{20}$ cm$^{-3}$ carriers estimated from the absorption coefficient $\alpha = 1.2 \times 10^5$ cm$^{-1}$ in Si. This is within an order of magnitude for the critical density for screening of the carrier-phonon interaction in Si [33]. In the case of GaAs, the photo-excitation of electrons above the band-gap causes saturation and/or permanent damage in GaAs surface at average pump powers exceeding 10 mW (0.17 mJ/cm$^2$). Therefore, all measurements in GaAs were carried out at or below 10 mW, corresponding to maximum of $N \approx 3.7 \times 10^{20}$ cm$^{-3}$ carriers based on the absorption coefficient $\alpha = 6.7 \times 10^5$ cm$^{-1}$ in GaAs at 397 nm. After reflecting from the sample, the probe was analyzed into polarization components parallel and perpendicular to that of the pump and each was detected with a photodiode. The resulting photocurrents were subtracted and after amplification their difference [$\Delta R_{eo}/R_0 = (\Delta R_k - \Delta R_l)/R_0 = (|E_k|^2 - |E_l|^2)/|E_0|^2$] was recorded versus the pump-probe delay, as shown in Fig. 2, where $E_k$ and $E_l$ are the components of the reflected probe light. The delay was scanned over 10 ps and averaged for 10,000 scans by using a shaker with 20 Hz-frequency. In order to examine the resonant behaviour in Si, the laser wavelength was varied between 2.99 eV (415 nm) and 3.16 eV (392 nm), which is the maximum tuning rage obtained with our 10 fs laser pulse by adjusting the phase matching angle of the 50 µm-thick BBO crystal for second harmonic generation of the laser fundamental.

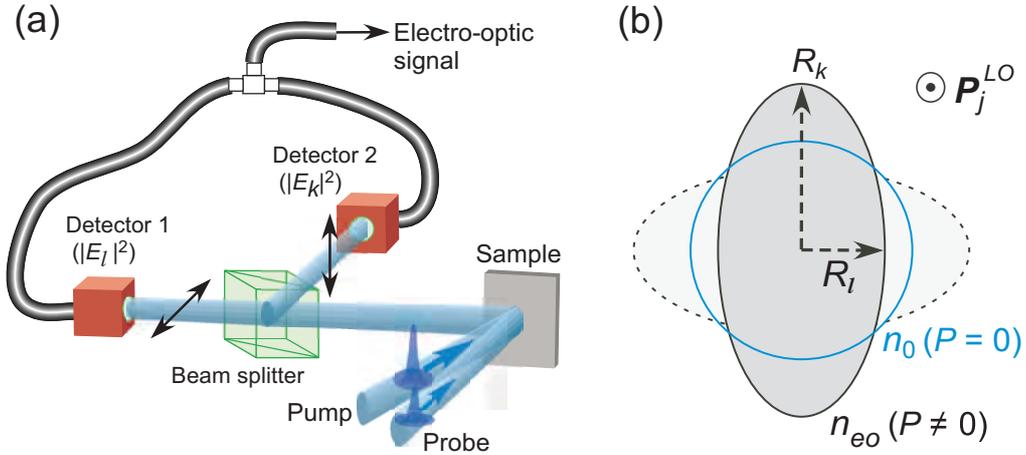

**Figure 2.** (a) Electro-optic detection of the anisotropic change in the refractive index. The incident angles of the pump and probe beams were nearly surface normal with <5º relative angle. The polarization of the probe was 45º with respect to the optical plane. (b) Isotropic refractive index ($n_0$) before the photoexcitation ($P$=0) and the anisotropic refractive index ($n_{eo}$) after the photoexcitation ($P \neq 0$). $\boldsymbol{P}_j^{LO}$ represent the longitudinal polarization induced by coherent LO phonon.

## 3. Results and analysis

### 3.1. Si

Figure 3 shows the transient electro-optic sampling signal recorded for different excitation photon energies between 2.99 - 3.16 eV with a constant pump power of 20 mW. For the 2.99 eV excitation, an aperiodic electronic response near zero delay dominates the signal. For higher energies, in addition, there appears a coherent oscillation with a period of ~64 fs that persists for ~10 ps due to the $k = 0$ coherent LO phonons [5, 6]. The phonon amplitude monotonically increases reaching maximum of $\Delta R_{eo}/R_0 \sim 2.5 \times 10^{-5}$ at the high-energy limit of the tuning range (Fig. 3 inset), where it is comparable to the electronic response. As with the spontaneous Raman spectra, the LO phonon signal is enhanced by resonance with the direct band gap of Si [32]. To support this conclusion, the resonant behaviour in inset of Fig. 3 is modelled with the imaginary part of the susceptibility $|\chi_2|^2$ [32]. This resonance enhancement explains the markedly more efficient coherent LO phonon excitation in Si than at 800 nm.

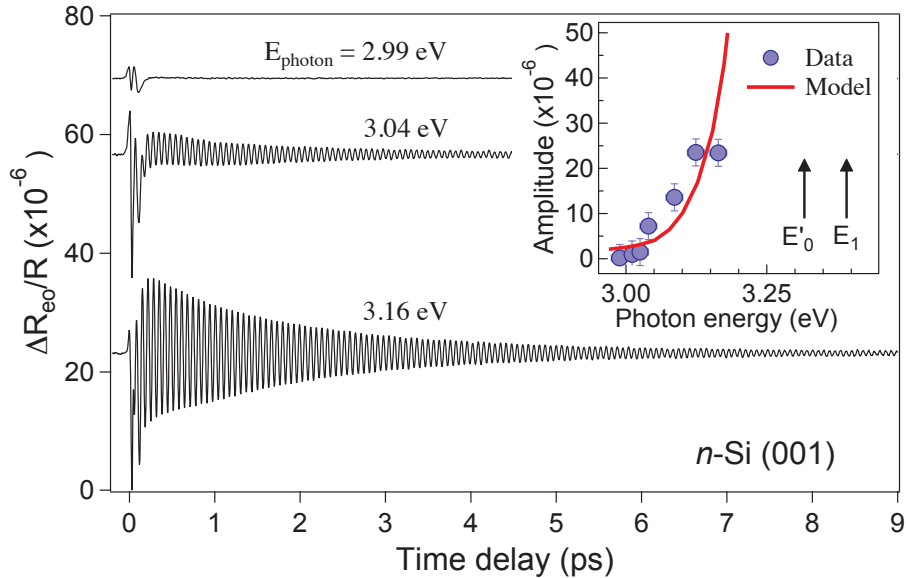

**Figure 3.** Transient e-o reflectivity obtained for different excitation energies in *n*-Si. The inset shows the electro-optic amplitude of the coherent LO phonon as a function of the photon energy. The solid curve is a fit with the model described in Ref. [32].

Figure 4 shows the analysis of the coherent LO phonon signal for different laser fluences corresponding to the initial photoexcited carrier densities in the $N \sim 3.3 \times 10^{19} - 1.0 \times 10^{20}$ cm$^{-3}$ range for the excitation and probing at 397 nm, which generates the maximum amplitude of the coherent LO phonon. Remarkably, fitting the phonon oscillations in short (~ 3 ps) segments to an exponentially damped cosine function, as we have done for Si in Ref. [34], indicates that both the dephasing time and the frequency change approximately linearly with the time delay.

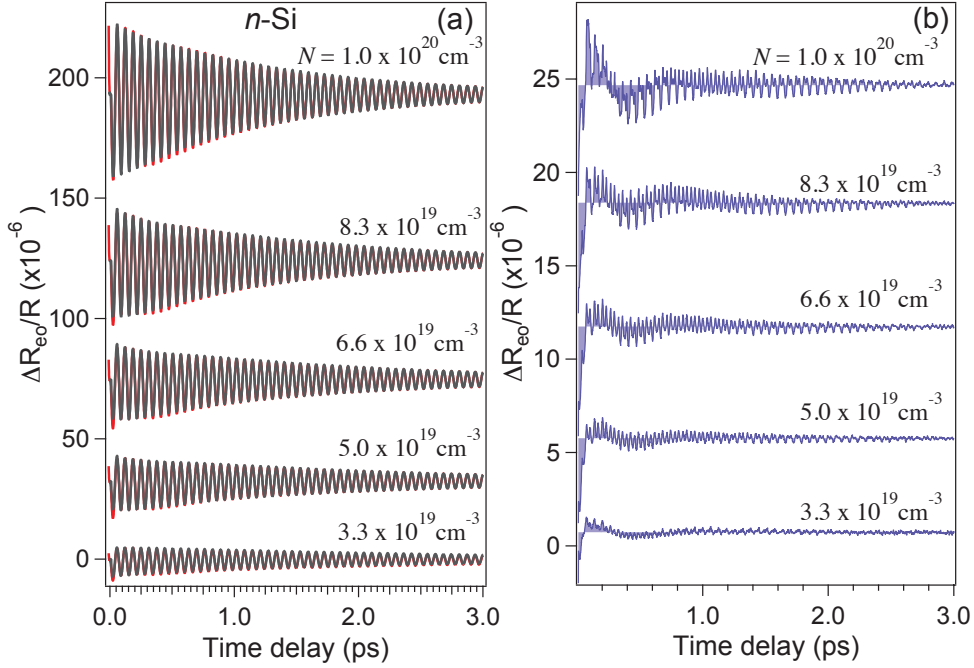

**Figure 4.** (a) Transient e-o reflectivity was observed with 397 nm (3.12 eV) excitation with various photoexcited carrier densities. The red lines are the experimental data and the black lines show the fit with equation (2). (b) The residual of the fit, showing the contributions from the fundamental and higher harmonics of the coherent LO phonon frequency.

As seen in Fig. 4(a), the $\Delta R_{eo}/R_0$ signal can be fit well to a damped oscillation with the frequency chirp ($\eta$) and the time-dependent relaxation time ($\tau_r + \delta t$),

$$F(t) = A\, exp\left[-t/(\tau_r + \delta t)\right] cos[(\omega_0 + \eta t)t + \phi], \qquad (2)$$

where $A$ is the amplitude, $\omega_0$ is the frequency, and $\phi$ is the initial phase of the coherent LO phonon. We obtained $\tau_r = 1.7$ ps, $\delta = 0.12$, $\omega_0 = 15.6$ THz, $\eta = 2 \times 10^{-2}$ ps$^{-2}$, and $\phi = 10$ degrees, at the photoexcited carrier density of $N \approx 1.0 \times 10^{20}$ cm$^{-3}$. The power, or more pertinently, the photoinduced carrier density dependence of the parameters obtained by fitting the data to Eq. (2) is presented in Ref. [34]. The frequency and dephasing time, *i.e.*, the phonon self-energy [35], vary approximately linearly with the carrier density. The coherent LO phonon dephasing by the anharmonic coupling, which in semimetals gives rise to dependence of the phonon frequency on the square of the amplitude [36], appears to be insignificant in Si. A finding uncovered from the residual of the fit, shown in Fig. 4(b), is that higher-order harmonics of the LO frequency with periods of 32 and 16 fs, respectively, also contribute to the optical response of the Si sample. It is interesting to note that the oscillatory phase of the 1st-order oscillation is found to be sine-like ($\phi \approx$ 60–90 degrees) at lower pump fluences than 40 mW (0.67 mJ/cm$^2$ or $N \approx 6.6 \times 10^{19}$ cm$^{-3}$), while at higher fluences ( $\geq$ 45 mW or 0.75 mJ/cm$^2$) it is cosine-like ($\phi = 10$ degrees) [16, 37].

To further investigate the character of the higher harmonics produced by the coherent LO phonon modulation, we obtain the Fourier transformed (FT) spectra, as shown in Fig. 5, which reveal a comb of frequencies corresponding to the fundamental LO phonon and its higher-harmonics. The harmonics appear exactly at the integer multiples of the fundamental frequency (15.6 THz), *i.e.*, at 31.2, 46.8, 62.4, 78.0, 93.6 and 109.2 THz, up to the seventh-harmonic at the highest fluence corresponding to $N \approx 1.0 \times 10^{20}$ cm$^{-3}$. From these results, we can exclude the higher-order Raman scattering because Raman overtones, appear as broad density-of-states features with a maxima around, but not exactly at the integer multiple of the fundamental frequency, and depend on the anharmonicity of the oscillator rather than the pump laser fluence [32].

As the pump fluence decreases the relative amplitudes of the higher-order harmonics decreases and only the 1st and 2nd orders are visible at the $N \approx 3.3 \times 10^{19}$ cm$^{-3}$ excitation. At $N \leq 1.7 \times 10^{19}$ cm$^{-3}$, moreover, only the fundamental response is detected (not shown). These data suggest that the modulation amplitude, $\propto A$, is smaller than that required for generation of the higher-orders or the dynamic range of our data acquisition system is insufficient for the observation of the higher-orders at the lower photoexcited carrier densities.

The FT spectra in Fig. 5 show interesting features that are related to the mechanism of higher harmonic generation: (i) the second harmonic (31.2 THz) exhibits a derivative shape with narrow bandwidth ($\Delta\Gamma$) whereas for third and higher harmonics the shapes are more asymmetric and the bandwidths significantly broader; and (ii) there is an odd–even order harmonic intensity alternation with even harmonics being more intense than the odd. These features can be reproduced in a simulation as described in Sec. 4.

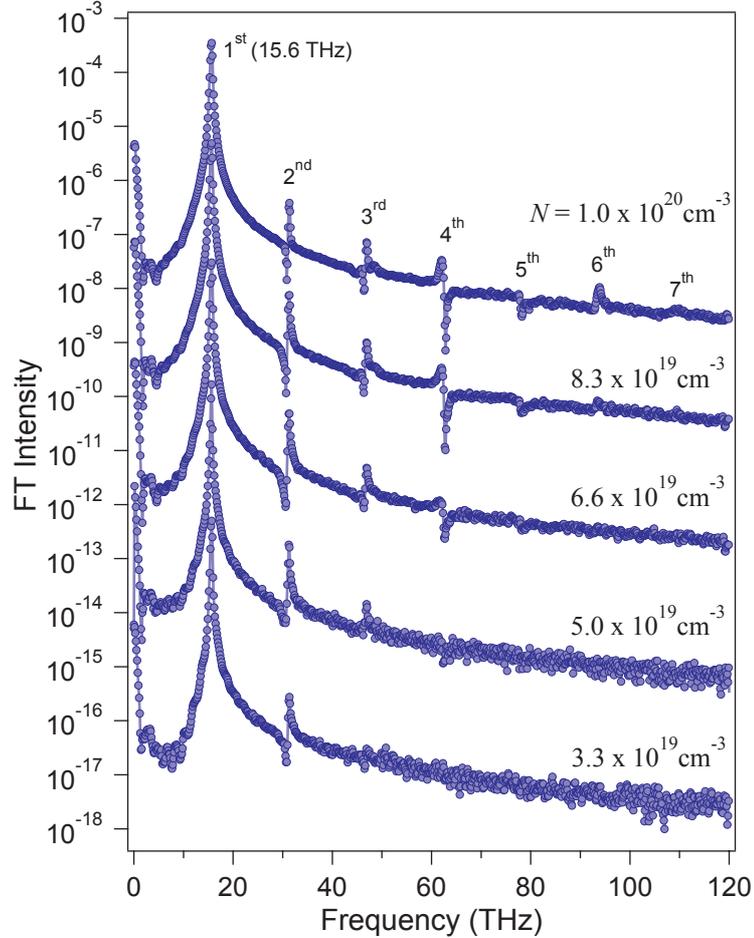

**Figure 5.** FT spectra obtained from the transient anisotropic reflectivity of *n*-Si at various photoexcited carrier densities. The spectra were offset in the vertical axis to distinguish each other.

### 3.2. GaAs

Figure 6 shows transient electro-optic sampling signal observed at different photoexcited carrier densities in *n*-GaAs(001). The amplitude of the coherent phonon oscillation increases with increasing the photoexcited carrier density. The time period of the coherent oscillation, which is roughly estimated to be ~127 fs, is significantly longer than that of the LO phonon period (114 fs) [3], implying that the dominant contribution to the coherent response in Fig. 6 is from the lower branch ($\omega_-$) of the LOPC modes [20, 38]. By fitting the time domain data to Eq. (2), we obtained $\tau_r = 0.8$ ps, $\delta \approx 0$, $\omega_0 = 7.8$ THz, $\eta = -3 \times 10^{-2}$ ps$^{-2}$, and $\phi = -4$ degrees, at the highest photoexcited carrier density of $N \approx 3.7 \times 10^{20}$ cm$^{-3}$. Note that the fit with only the LOPC contribution ignores the weak bare LO phonon contribution, as discussed in Ref. [38]. From the residual of the fit, shown in Fig. 6(b), the higher-order harmonics of the LOPC frequency with periods of 63 and 42 fs, respectively, also contribute to the optical response of *n*-GaAs.

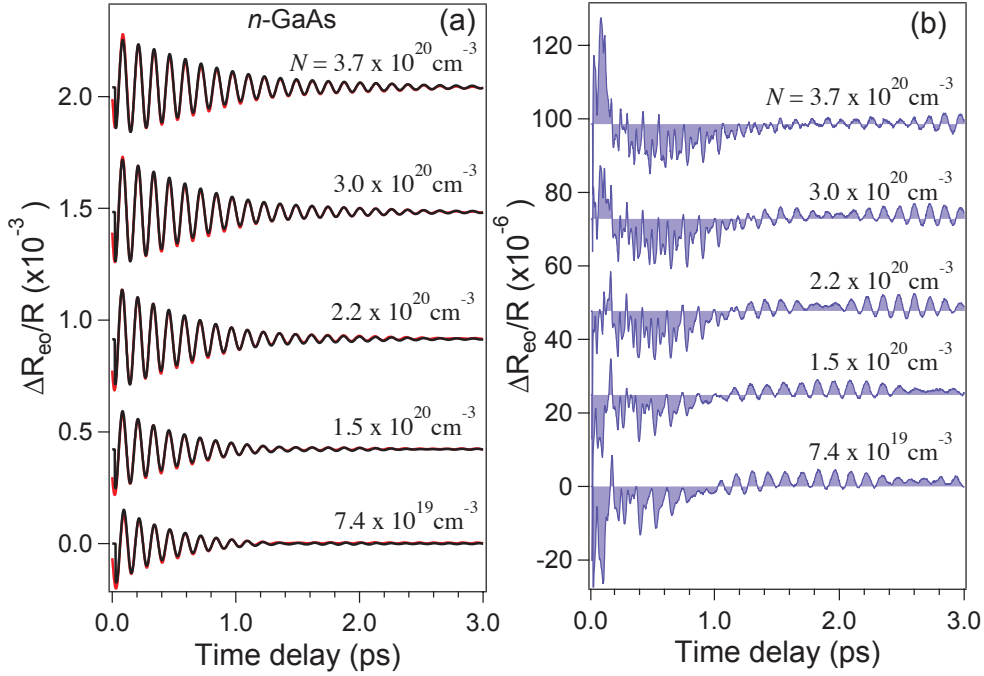

**Figure 6.** (a) Transient anisotropic reflectivity observed by 397 nm (3.12 eV) excitation with various photoexcited carrier densities in *n*-GaAs. The red line represents the experimental data and the black line shows the fit with Eq. (2). (b) The residual of the fit, showing the fundamental and higher harmonics of the coherent LOPC frequency.

Figure 7 summarizes the FT spectra of the transient electro-optic sampling signal observed at different photoexcited carrier densities in *n*-GaAs. There appears dominantly one peak at almost the same position of the TO phonon (7.8 THz). This peak very slightly shifts up toward the frequency of the bare LO phonon (8.8 THz) and its linewidth becomes broad as the carrier density $N$ decreases. For the interaction of electron plasma with the ionic lattice the lower branch of the LOPC modes ($\omega_-$) should approach the TO phonon frequency as the photoexcited carrier density increases [19–21]. On the contrary, we observe the LOPC mode decrease with the carrier density from the LO mode limit to the TO mode limit; this behaviour is characteristic of the interaction of the LO phonon with doped hole plasma [38]. Therefore, we assign the observed mode at 7.8 THz as the LOPC mode due to interaction of the lattice vibrations with the photoexcited holes. The FT spectra reveal a comb of frequencies corresponding to the fundamental LOPC mode and its higher-harmonics. The harmonics appear almost exactly at the multiple of the fundamental frequency (7.8 THz), *i.e.*, at 15.7, 23.5, 31.7, 39.3, 47.4 and 55.3 THz, up to the seventh-harmonic. The spectra show similar features to the case of Si, *i.e.*, (i) the second harmonic (15.7 THz) exhibits a derivative shape with narrow bandwidth ($\Delta\Gamma$) whereas for the third and higher harmonics the shapes are more asymmetric and bandwidth significantly broader; (ii) the harmonic intensities alternate with the odd harmonics being more intense than the even.

## 4. Model simulations
Our observation of the higher harmonics in Si and GaAs can be related to the modulation of index of refraction by the coherent LO phonon (LOPC mode in GaAs) and subsequent generation of collinear sidebands, similar to what is commonly found in transmission through transparent gas phase molecular systems [24, 25]; that is, coherent LO phonons driven by pump light modulate the amplitude and the phase of the probe electric field. To gain concrete evidence and insight into the observation mechanism of the phonon frequency comb, we carried out the modelling of transient reflectivity from semiconductor surfaces for resonant excitation.

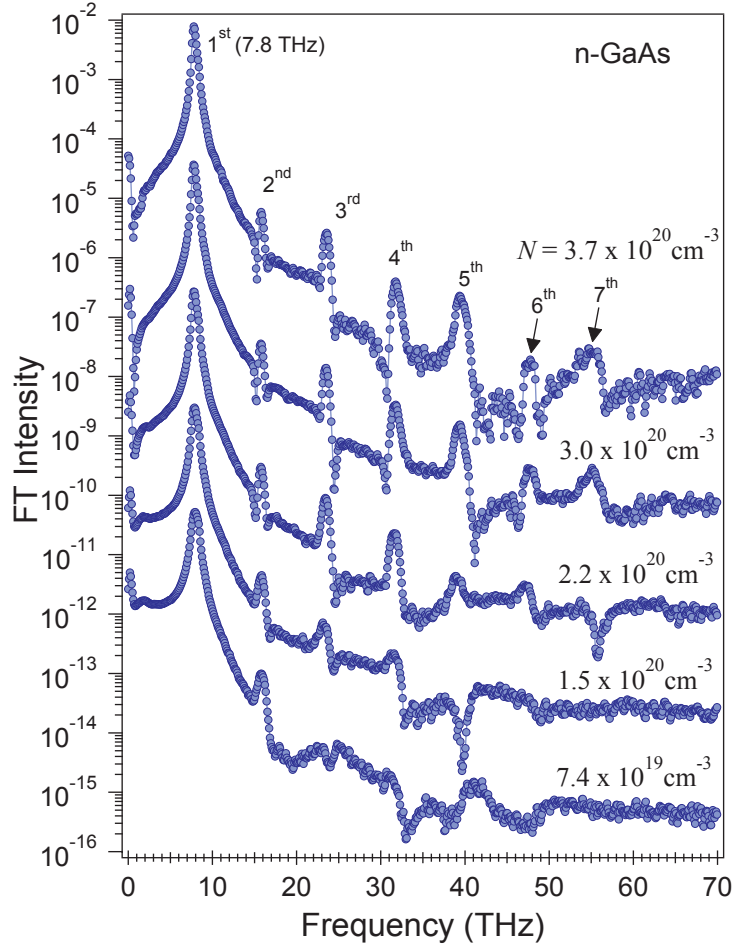

**Figure 7.** FT spectra obtained from the transient anisotropic reflectivity at various photoexcited carrier densities in *n*-GaAs. The spectra were offset in the vertical axis to distinguish each other.

The main idea of the generation of high harmonic frequency combs is the modulation of the probe light via the change in the index of refraction by coherent phonon oscillation [37]. This point of view is slightly different from the Raman scattering, where one observes the scattering light intensity rather than the change in the optical constant. The physical mechanism for the change in the real and imaginary parts of the index of refraction is the band gap renormalization, *i.e.*, the energy shift ($\Delta\varepsilon$) of the band edge by the optical deformation potential $\Xi$, which is defined by $\Delta\varepsilon = \Xi\left|\Delta Q\right|/Q_0$ [39], where $\Delta Q$ is the internal displacement (Si-Si) due to the optical phonon, and $Q_0$ is the static value of Si-Si bond length. The dependence of the band gap and therefore the optical constants is expected for the $\Gamma_{25'}$ symmetry vibration of a crystal with inversion symmetry. Although GaAs has zincblende rather than diamond structure and is a polar material, the physical processes responsible for the phonon frequency comb generation are exactly analogous. The electric field amplitude as a function of the time delay $t$ with an amplitude and phase modulation can be expressed as,

$$E_r(t) = \left\{a(t)\cos(\omega_0 t) + R_0\right\}\cos\left[\omega t - \delta(t)\cos(\omega_0 t + \beta)\right], \qquad (3)$$

where $a(t) = a_0 e^{-\gamma t}$ and $\delta(t) = \delta_0 e^{-\gamma t}$ are the magnitudes of the amplitude (AM) and the phase modulation (PM), $R_0$ is the static reflectivity, $\omega$ is a carrier frequency of the probe pulse, $\omega_0$ is the phonon frequency, and $\beta$ is the phase delay between the amplitude and phase modulations. The value of $\beta$ is expected to be small and would be nearly zero because both AM and PM originate from the same coherent phonon mode. Note that both $a(t)$ and $\delta(t)$ decay with the amplitude of the coherent optical phonon; $\gamma$ represents the damping rate of the coherent phonon, which is 0.6 ps$^{-1}$ for Si in our simulation. Using the Jacobi-Anger expansion, we can invoke

$$cos\big[\delta(t)cos\big(\omega_0 t + \beta\big)\big]$$
$$= J_0\big(\delta(t)\big) - 2J_2\big(\delta(t)\big)cos\big(2\omega_0 t + 2\beta\big) + 2J_4\big(\delta(t)\big)cos\big(4\omega_0 t + 4\beta\big) + \cdots,$$

$$sin\big[\delta(t)cos\big(\omega_0 t + \beta\big)\big]$$
$$= 2J_1\big(\delta(t)\big)cos\big(\omega_0 t + \beta\big) - 2J_3\big(\delta(t)\big)cos\big(3\omega_0 t + 3\beta\big) + 2J_5\big(\delta(t)\big)cos\big(5\omega_0 t + 5\beta\big) + \cdots,$$

where $J_n(z)$ is the Bessel function of the first kind on the order of $n$. Using these relations, Eq. (3) can now be rewritten as,

$$
\begin{aligned}
E_r(t) = &\big[a(t)cos\big(\omega_0 t\big) + R_0\big] \times \\
&\Big\{ J_0\big(\delta(t)\big)cos\big(\omega t\big) + J_1\big(\delta(t)\big)sin\big[\big(\omega + \omega_0\big)t + \beta\big] + J_1\big(\delta(t)\big)sin\big[\big(\omega - \omega_0\big)t - \beta\big] \\
&- J_2\big(\delta(t)\big)cos\big[\big(\omega + 2\omega_0\big)t + 2\beta\big] - J_2\big(\delta(t)\big)cos\big[\big(\omega - 2\omega_0\big)t - 2\beta\big] \\
&- J_3\big(\delta(t)\big)sin\big[\big(\omega + 3\omega_0\big)t + 3\beta\big] - J_3\big(\delta(t)\big)sin\big[\big(\omega - 3\omega_0\big)t - 3\beta\big] \\
&+ J_4\big(\delta(t)\big)cos\big[\big(\omega + 4\omega_0\big)t + 4\beta\big] + J_4\big(\delta(t)\big)cos\big[\big(\omega - 4\omega_0\big)t - 4\beta\big] \\
&+ \cdots \Big\}.
\end{aligned}
\tag{4}
$$

The second part in Eq. (4), { }, indicates the phonon sidebands are generated due to the phase modulation.

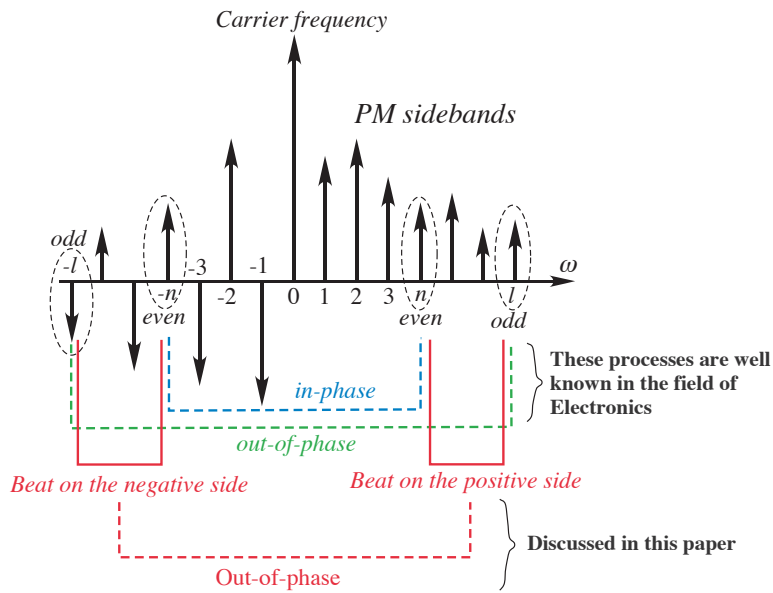

**Figure 8.** The sideband generation via the phase modulation (PM).

It should be noted that phase modulation itself generates only the even-harmonic-orders [37], as discussed in the following. The phase modulation can be written by,

$$e^{i\{\omega t - \delta \cos(\omega_0 t + \beta)\}} = e^{i\omega t} e^{-i\delta \cos(\omega_0 t + \beta)} = e^{i\omega t} \sum_{-\infty}^{\infty} (-i)^n J_n(\delta) e^{-in(\omega_0 t + \beta)}, \quad (5)$$

where each term in the summation corresponds to the complex amplitude of phonon sideband. The phonon sidebands with the orders up to $\pm \delta$ would appear in frequency domain. Since our observation corresponds to the phonon sideband beat, we calculate; $\left| e^{i\{\omega t - \delta \cos(\omega_0 t + \beta)\}} \right|^2$ and its Fourier transform, whose spectral components become;

$$\left| e^{i\omega t} \sum_{-\infty}^{\infty} (-i)^n J_n(\delta) e^{-in(\omega_0 t + \beta)} \right|^2. \quad (6)$$

When we expand these components, odd-terms of the phonon sideband beat become;

$$(-i)^n (i)^l J_n(\delta) J_l(\delta) e^{-in(\omega_0 t + \beta)} e^{il(\omega_0 t + \beta)}, \quad (|l - n| \text{ is an odd number}) \quad (7)$$

which is finally found to vanish. Here the relative phase between each term is defined as $(-i)^n (i)^l$ in Eq. (7). Examining the expansion of the odd-terms, we find that the beating between the phonon sidebands in negative frequency region (the beating between $-n$ and $-l$ orders) has the opposite phase from that in the positive region (the beating between $+n$ and $+l$ orders; see Fig. 8). For example, in the case of $|l - n|$ being odd number, $(-i)^n (i)^l$ with $n=1$, $l=4$ leads $(-i)^1 (i)^4 = -i$ and $(-i)^n (i)^l$ with $n = -1$, $l = -4$ leads $(-i)^{-1}(i)^{-4} = +i$, indicating the opposite phase. By contrast, in the case of $|l - n|$ being even number, $(-i)^n (i)^l$ with $n=1$, $l=3$ leads $(-i)^1 (i)^4 = -i$ and $(-i)^n (i)^l$ with $n= -1$, $l= -3$ leads $(-i)^{-1}(i)^{-3} = -1$, indicating the same phase.

Thus, the relative phase between the phonon sideband beating is opposite between the positive ($\omega > 0$) and negative frequency side ($\omega < 0$). Consequently, the beating disappears when $|l - n|$ is an odd number. On the contrary, the relative phase becomes the same on the positive and negative frequency sides when $|l - n|$ is an even number, therefore the beating survives when $|l - n|$ is an even number.

In our model the pump pulse generates coherent optical phonons, which modulate the index of the refraction at the phonon frequency $\omega_0$. The amplitude and the phase of the probe pulse are modulated by the coherent phonon-induced index modulation [Eq. (3)]. The phase modulation is possible because the probe beam penetrates into the sample (optical penetration depth ~82 nm for Si and ~15 nm for GaAs) in the process of being reflected. The oscillatory observable in the electro-optic sampling measurements is the longitudinal phononic part of the difference between $\left| E_k(t) \right|^2$ and $\left| E_l(t) \right|^2$ components, which is expressed by $\left| E_{\text{Pr}}(t) \right|^2 / E_0^2$ ;

$$\left| E_{\text{Pr}}(t) \right|^2 / E_0^2 \approx 2a(t) R_0 \left( \frac{1}{2} J_0^2 + \frac{3}{2} J_1^2 + J_2^2 + J_3^2 + J_4^2 - J_0 J_2 - J_1 J_3 - J_2 J_4 \right) \cos(\omega_0 t)$$

$$+ 2R_0^2 \left( \frac{1}{2} J_1^2 - J_0 J_2 - J_1 J_3 - J_2 J_4 \right) \cos(2\omega_0 t) + \frac{1}{4} a(t)^2 J_0^2 \cos(2\omega_0 t)$$

$$+2a(t)R_0\left(\frac{1}{2}J_1^2+\frac{1}{2}J_2^2-J_0\,J_2-2J_1\,J_3+J_0\,J_4-J_2\,J_4\right)cos(3\omega_0 t)$$

$$+2R_0^2\left(\frac{1}{2}J_2^2-J_1\,J_3+J_0\,J_4\right)cos(4\omega_0 t)$$

$$+2a(t)R_0\left(\frac{1}{2}J_2^2+\frac{1}{2}J_3^2-J_1\,J_3+J_0\,J_4-J_2\,J_4\right)cos(5\omega_0 t)$$

$$+2R_0^2\left(\frac{1}{2}J_3^2-J_2\,J_4\right)cos(6\omega_0 t)$$

$$+2a(t)R_0\left(\frac{1}{2}J_3^2+\frac{1}{2}J_4^2-J_2\,J_4\right)cos(7\omega_0 t)+\cdots. \qquad (8)$$

Here we neglect the terms higher than the 5th order in Eq. (4) because of the limited bandwidth of our laser (~100 THz). Note that including 5th and 6th order terms in Eq. (4), predicts responses up to 12th order (~200 THz), which is beyond the bandwidth of our laser and cannot be observed. It should be also noted that, the amplitude of the odd-orders has a contribution from the '$a(t)$' term whereas the even-orders do not; this explains the odd–even harmonic intensity alternation.

The simulated time-domain signal and FT spectra are displayed in Fig. 9. These calculations are based on the amplitude and phase modulations of laser electric field via coherent phonon oscillation. Overall, the appearance of the higher harmonics is consistent with the experiments, including: (i) bandwidth broadening as the order number increases; (ii) the odd–even-order intensity alternation with even harmonics being more intense than the odd; and (iii) the alternating asymmetric lineshapes of the higher orders.

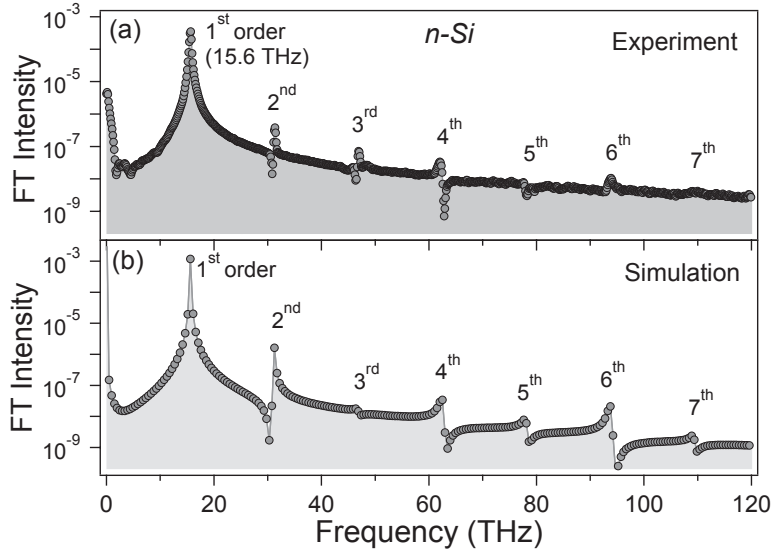

**Figure 9.** FT spectrum obtained from the transient anisotropic reflectivity in *n*-Si at $N \approx 1.0 \times 10^{20}$ cm$^{-3}$ (upper), comparing with the simulation (bottom).

The range of values for the band gap renormalization, *i.e.*, the energy shift ($\Delta\varepsilon$), is $\Delta\varepsilon = 0.1 - 0.2$ eV for the photoexcited carrier density of $N \approx 1.0 \times 10^{21}$ cm$^{-3}$ based on calculations for GaAs and Si [40]. Therefore, the photoexcited carrier density tunes Si into resonance with the excitation laser. A manifestation of the band gap renormalization in Si is the dependence of the coherent phonon phase $\phi$ on the excitation density. At low densities the coherent phonon phase $\phi$ is close to the impulsive limit ($\phi = 90°$), indicating that the applied force is dominated by the Raman susceptibility [13, 16].

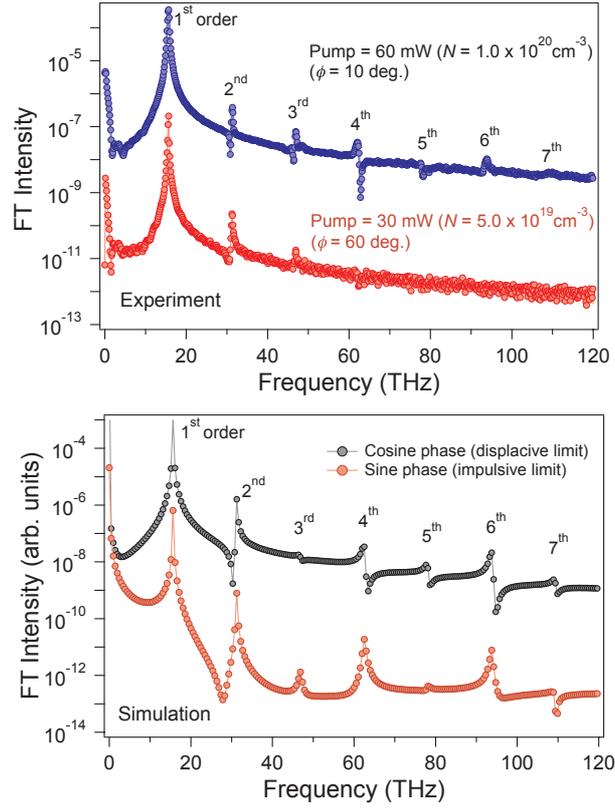

**Figure 10.** (a) Experimental FT spectrum obtained at the two different fluences in *n*-Si. (b) The simulation results for different phases of the coherent LO phonon.

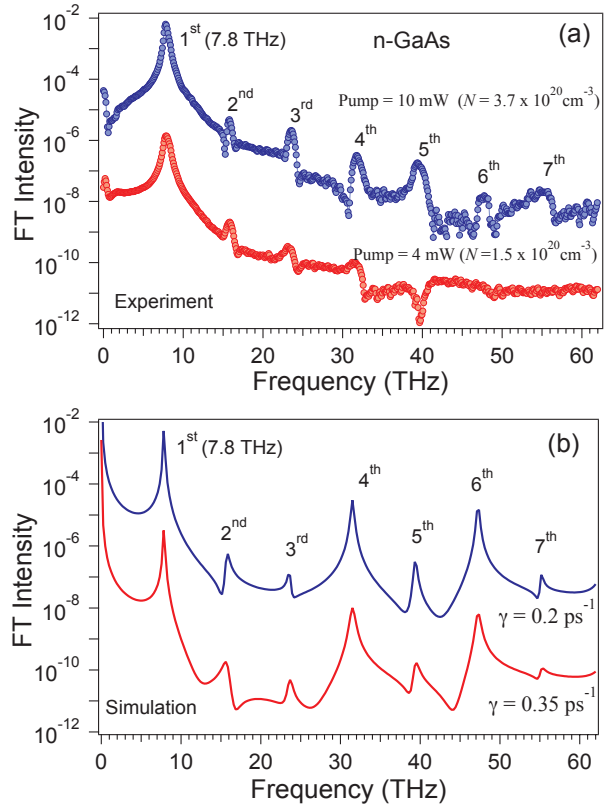

**Figure 11.** (a) Experimental FT spectrum obtained at the two different fluences in *n*-GaAs. (b) The simulation results for the two different damping γ of the LOPC mode.

By contrast, at high densities the band gap renormalization during the excitation brings the direct band gap of Si into resonance with the excitation light because the anisotropic excitation of L-valley carriers exerts a displacive electrostrictive force ($\phi = 0°$) [7, 16]. We note that when replacing the AM and PM by sine functions in equation (3), as appropriate for the impulsive limit, the simulation reproduces the FT spectra obtained with 30 mW excitation corresponding to a lower photoexcited carrier density ($N \approx 0.5 \times 10^{20}$ cm$^{-3}$), as shown in Fig. 10.

As shown in Fig. 11, in *n*-GaAs the simulations reproduce the overall lineshape well, although the peak positions and the asymmetries in the experiment are not simple because of the nature of the LOPC modes, which is due to coupling between the LO phonon and spatially separated hole and electron plasmas [38]. As seen in the time domain signal the dephasing time of the LOPC mode depends on the pump fluence, and becomes shorter at higher fluences. These changes result in the modification of the lineshapes at low fluences. There, the overall lineshape can be reproduced by taking a larger value for the damping, as shown in Fig. 11(b).

## 5. Summary

In summary, we have generated and observed the coherent phonon frequency combs in reflection from *n*-Si and *n*-GaAs surfaces by using ~10 fs laser pulses operated in near-UV region. The transient change in the reflectivity measured by a fast-scan pump-probe technique in the electro-optic mode revealed the comb harmonics up to 7th-order (109.2 THz) of the coherent LO phonon in the case of Si. In the case of GaAs we observe the harmonics of the LOPC mode up to 7th-order (55.2 THz). In both cases, the coherent longitudinal polarization generated by the resonant or near-resonant excitation of the direct band gap, modulate the index of refraction of the samples in time and frequency domains. Our results demonstrate the possibility of a semiconductor based ultra-broadband THz phononic modulator for the manipulation of the phase and amplitude of laser pulses [41], whose wavelength range spreads from 50 nm (x-ray) [42, 43] to 1 mm (THz-ray), if we introduce larger phonon displacement to modulate x-rays and infrared phonon absorption to modulate THz-rays.

### Acknowledgements

This work was supported in part by NSF under grants CHE-0650756 and CHE-0911456.